\documentclass[aps,prl,reprint,groupedaddress,showpacs,nofootinbib]{revtex4-1}
\usepackage{graphicx}
\usepackage{amsmath}
\usepackage{amssymb}
\usepackage{hyperref}

\hypersetup{
    colorlinks=true,
    linkcolor=blue,
    citecolor=blue,
    filecolor=blue,
    urlcolor=blue,
}

\begin{document}

\textbf{Comment on ``Self-Averaging Stochastic Kohn-Sham Density-Functional Theory''}

\

In a recent Letter \cite{SKS}, Baer \textit{et al.} present a stochastic method
 for Kohn-Sham density functional theory (DFT) calculations.
They converge the total energy per electron, $E/N$,
 to determine the number of statistical samples, $s$.
Self-averaging of $E/N$ allows it to achieve constant error
 while reducing $s$ with increasing $N$.
However, when used as a convergence criterion, $E/N$
 does \textit{not} guarantee the convergence
 of quantities that do not self-average.
Their errors relative to $E/N$ grow with $N$ until saturating at a large maximum error.
This includes the electron density, atomic forces, and orbital energies,
 which the authors claim can be calculated reliably.
When converging $E/N$, 
 computational costs exhibit novel sublinear scaling in
 $N$ as $s\rightarrow1$, beyond which the cost scales linearly.
 If any non-self-averaging values are converged,
 sublinear scaling ceases and the cost prefactor increases significantly.

The Letter calculates the stochastic expectation value of $\mathbf{X}$ for $N$ electrons on $M$ grid points with $s$ samples as
\begin{equation}
 \mathrm{Tr}_s( \boldsymbol{\rho} \mathbf{X} ) \approx (M/s) \sum\nolimits_{i=1}^s \mathbf{v}_i^\dag \sqrt{\boldsymbol{\rho}}_q \mathbf{X} \sqrt{\boldsymbol{\rho}}_q \mathbf{v}_i ,
\end{equation}
 where $\mathbf{v}_i$ is a random unit vector,
 $\boldsymbol{\rho}$ is a density matrix,
 $\sqrt{\boldsymbol{\rho}}_q$
 is a polynomial approximation of $\tfrac{1}{2} \mathrm{erfc}[\beta (\mathbf{H}-E_F)]$
 with degree $q$, $\mathbf{H}$ is the Hamiltonian, and $E_F$ is the Fermi energy.
$\mathbf{v}_i$ are sampled from the random-phase ensemble
 of vectors  \cite{randomphase}, which converges the standard error as
\begin{align}
 \epsilon_s(\boldsymbol{\rho} \mathbf{X}) &= | \mathrm{Tr}_s(\boldsymbol{\rho} \mathbf{X}) - \mathrm{Tr}(\boldsymbol{\rho} \mathbf{X}) | \approx \sigma(\boldsymbol{\rho} \mathbf{X})/\sqrt{s}, \notag \\
 \sigma(\boldsymbol{\rho} \mathbf{X})^2 &= \sum\nolimits_i \left( |\lambda_i(\sqrt{\boldsymbol{\rho}} \mathbf{X} \sqrt{\boldsymbol{\rho}})|^2 - | [\sqrt{\boldsymbol{\rho}} \mathbf{X} \sqrt{\boldsymbol{\rho}}]_{ii} |^2 \right) , \label{random_phase}
\end{align}
 where $\lambda_i(\sqrt{\boldsymbol{\rho}} \mathbf{X} \sqrt{\boldsymbol{\rho}})$
 are the eigenvalues of $\sqrt{\boldsymbol{\rho}} \mathbf{X} \sqrt{\boldsymbol{\rho}}$.
 
From Eq. (\ref{random_phase}) it is evident that expectation values with similar magnitudes,
 $\mathrm{Tr}(\boldsymbol{\rho} \mathbf{X}) \sim \mathrm{Tr}(\boldsymbol{\rho} \mathbf{Y})$,
 will \textit{not} always have similar standard errors,
 $\epsilon_s(\boldsymbol{\rho} \mathbf{X}) \sim \epsilon_s(\boldsymbol{\rho} \mathbf{Y})$,
 if the spectra of $\mathbf{X}$ and $\mathbf{Y}$ differ.
For $E/N = \mathrm{Tr}(\boldsymbol{\rho}\mathbf{H}/N) \propto 1$, $\sigma(\boldsymbol{\rho}\mathbf{H}/N) \propto 1/\sqrt{N}$
 from $\propto N$ eigenvalues of magnitude $\propto 1/N$.
Self-averaging is the reduction of $\sigma$ with $N$.
For other $\mathrm{Tr}(\boldsymbol{\rho}\mathbf{X}) \propto 1$, $\sigma(\boldsymbol{\rho}\mathbf{X}) \propto 1$
 from $\propto 1$ eigenvalues of magnitude $\propto 1$.
They do not self-average,
 and converged $E/N$ does not imply their convergence for $N \gg 1$.
The spatial structure of eigenvectors is irrelevant here.
They are localized for the electron density and delocalized for orbital energies,
 but $\sigma \propto 1$ in either case.

\begin{figure}
\includegraphics{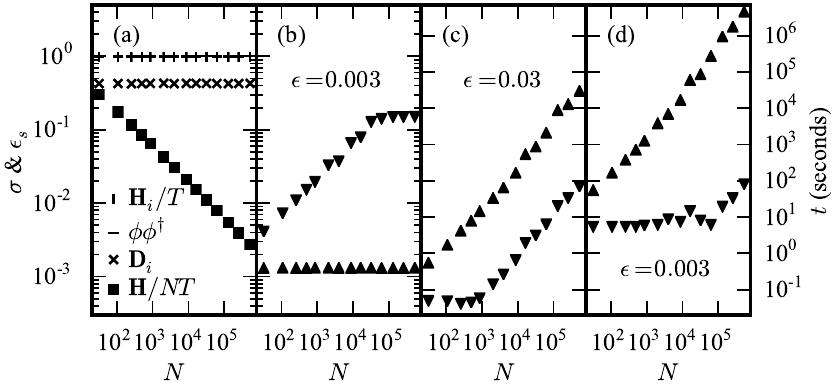}
\caption{\label{cube_sim} (a) example $\sigma$ values, (b) electron density errors, and (c,d) runtimes $t$ \cite{supplement}.
 In (b-d), we set $s$ to converge either $\epsilon_s(\mathbf{H}/NT)$ ($\blacktriangledown$)
 or all four $\epsilon_s$ values ($\blacktriangle$) to a target standard error $\epsilon$.
 All tests use $q=600$, $\beta T = 30$, and $s_{\min} = 8$.}
\end{figure}

We examine the predicted $\sigma$ scaling on a simple-cubic tight-binding model of cubic nanoparticles with $2N$ sites
 and hopping energy $T$ in Fig. \ref{cube_sim}.
Dimensionless values of $\sigma$ for $\mathbf{H}/NT$,
 a local energy density $\mathbf{H}_i/T$ ($\mathbf{H}=\sum_i{\mathbf{H}_i}$, $\mathbf{H}_i$ is hopping to and from site $i$),
 an electron density $\mathbf{D}_i$, and an ionization energy $\boldsymbol{\phi}\boldsymbol{\phi}^\dag$
  ($\boldsymbol{\phi}$ is the highest occupied orbital) all match predictions.
$E/N$ shows self-averaging, $\sigma \propto 1/\sqrt{N}$, but the other quantities do not, $\sigma \propto 1$.
The error in electron density increases with $N$ until a $\approx 0.2$ maximum
 using the $E/N$ convergence criterion.

Based on runtimes in Fig. \ref{cube_sim}, $s\propto\sigma^2/\epsilon^2$, and $t \propto N$ per sample,
 convergence uses $t \propto \max\{\epsilon^{-2},N\}$ for $E/N$ and $t \propto \epsilon^{-2}N$ for all values.
With all values converged,
 the sublinear-to-linear crossover at $s\propto1$ and $N\propto\epsilon^{-2}$ for $E/N$
 vanishes and the cost prefactor grows by $\epsilon^{-2}$.

Electron density errors cause errors in the Hartree and exchange-correlation potentials,
 which bias other values.
For runtimes in the Letter, we estimate the increased cost of converging the electron density.
Density errors are not reported directly, but cause force errors of $\approx 1$ eV/\AA \ for $s\approx10^3$.
$0.05$ eV/\AA \ is a representative stochastic force error in the literature \cite{QMC_force},
 which will require $s \approx4\times10^5$ here.
Runtimes in Fig. 2 of the Letter are $\approx 2 \times 10^{-3} N s$ hours (at $s \approx 2 \times 10^4 / \sqrt{N}$)
 for the stochastic calculations and $\approx 10^{-7} N^3$ hours for the conventional calculations.
Using the estimated value for $s$, the stochastic method becomes faster
 at $N\approx$ 90,000 rather than $N\approx$ 3,000.

Stochastic quantum Monte Carlo methods use a local energy density as a convergence criterion \cite{QMC_converge}.
It does \textit{not} self-average, which avoids the errors discussed here.

\begin{acknowledgments}
Sandia National Laboratories is a multi-program laboratory managed and  
operated by Sandia Corporation, a wholly owned subsidiary of Lockheed 
Martin Corporation, for the U.S. Department of Energy's National  
Nuclear Security Administration under contract DE-AC04-94AL85000.
\end{acknowledgments}

Jonathan E. Moussa$^*$ and Andrew D. Baczewski \\
\footnotesize Sandia National Laboratories, Albuquerque, NM 87185, USA \\
$^*$godotalgorithm@gmail.com


\begin{thebibliography}{9}
\bibitem{SKS} R. Baer, D. Neuhauser, and E. Rabani, \href{http://dx.doi.org/10.1103/PhysRevLett.111.106402}{Phys. Rev. Lett.} \textbf{111}, 106402 (2013).
\bibitem{randomphase} T. Iitaka and T. Ebisuzaki, \href{http://dx.doi.org/10.1103/PhysRevE.69.057701}{Phys. Rev. E} \textbf{69}, 057701 (2004).
\bibitem{supplement} See arXiv source files for implementation details.
\bibitem{QMC_force} A. Badinski and R. J. Needs, \href{http://dx.doi.org/10.1103/PhysRevE.76.036707}{Phys. Rev. E} \textbf{76}, 036707 (2007).
\bibitem{QMC_converge} P. J. Reynolds, D. M. Ceperley, B. J. Alder, and W. A. Lester Jr., \href{http://dx.doi.org/10.1063/1.443766}{J. Chem. Phys.} \textbf{77}, 5593 (1982).
\end{thebibliography}
\end{document}